\documentclass[a4paper,onecolumn]{agn6}
\usepackage{natbib}
\usepackage{graphicx}
\usepackage[a4paper]{hyperref}
\idline{}{1}
\begin{document}
   \title{Overdensity of X-Ray sources near 3C 295: a candidate filament
}

   \author{V. D'Elia, F. Fiore \and F. Cocchia }

   \institute{INAF - Osservatorio Astronomico di Roma
via di Frascati 33, 00040 Monteporzio Catone (RM) Italy  
\email{delia@mporzio.astro.it} 
             }





   \abstract{
We present a statistical analysis of the Chandra observation of 
the source field around the 3C 295 galaxy cluster ($z=0.46$) 
aimed at the search for clustering of X-ray sources.   
Three different methods of analysis, namely a chip by chip 
logN-logS, a two dimensional Kolmogorov-Smirnov test, and the angular
correlation function (ACF) show a strong overdensity of
sources in the North-East of the field. In particular, the ACF shows a clear 
signal on scales of $0.5\div 5$ arcmin,
correlation angle in the $0.5-7$ keV band, $\theta_0=8.5^{+6.5}_{-4.5}$, $90$\% 
confidence limit (assuming a power law ACF with
slope $\gamma=1.8$). 
This correlation angle is $> 2$ times higher than that 
of a sample of $8$ ACIS-I field at the $2.5 \; \sigma$ confidence level.
If this overdensity is spatially associated to
the  cluster, we are observing a 'filament' of the large scale structure of the Universe.
We discuss some first results that seem to indicate such an association.  

   }
   \authorrunning{V.D'Elia et al.}
   \titlerunning{X-Ray Sources Near 3C 295}
   \maketitle


%

\section{Introduction}


N-body and hydrodynamical simulations show that high redshift
clusters of galaxies lie at the nexus of several filaments of galaxies
(see e.g. Peacock 1999). Such filaments map out the ``cosmic web'' 
of voids and filaments of the large scale structure of the Universe. Thus, rich
clusters represent good indicators of regions of sky where several
filaments converge. The filaments themselves could be mapped out by
Active Galactic Nuclei (AGNs), assuming that AGNs trace galaxies. 

Cappi et al. (2001) studied the distribution of the X-ray sources
around 3C 295 ($z=0.46$). The cluster was observed with the ACIS-S CCD
array for a short exposure time ($18$ ks). 
They  found a high source surface density in the $0.5-2$ keV band which
exceeds the ROSAT (Hasinger et al. 1998) and Chandra (Giacconi et
al. 2002) logN-logS by a factor of $2$,  with a significance of $2\sigma$ . 

In this work an analysis of a deeper  ($92$ ks) Chandra observation of
3C 295 is performed, to check whether the overdensity of such a field
in the $0.5-2$ keV band is real or not, to define the structure of the
overdensity, and to extend the above considerations to the $2-10$ keV
band.

%
   \begin{figure}
   \centering
\resizebox{\hsize}{!}{\includegraphics[clip=true]{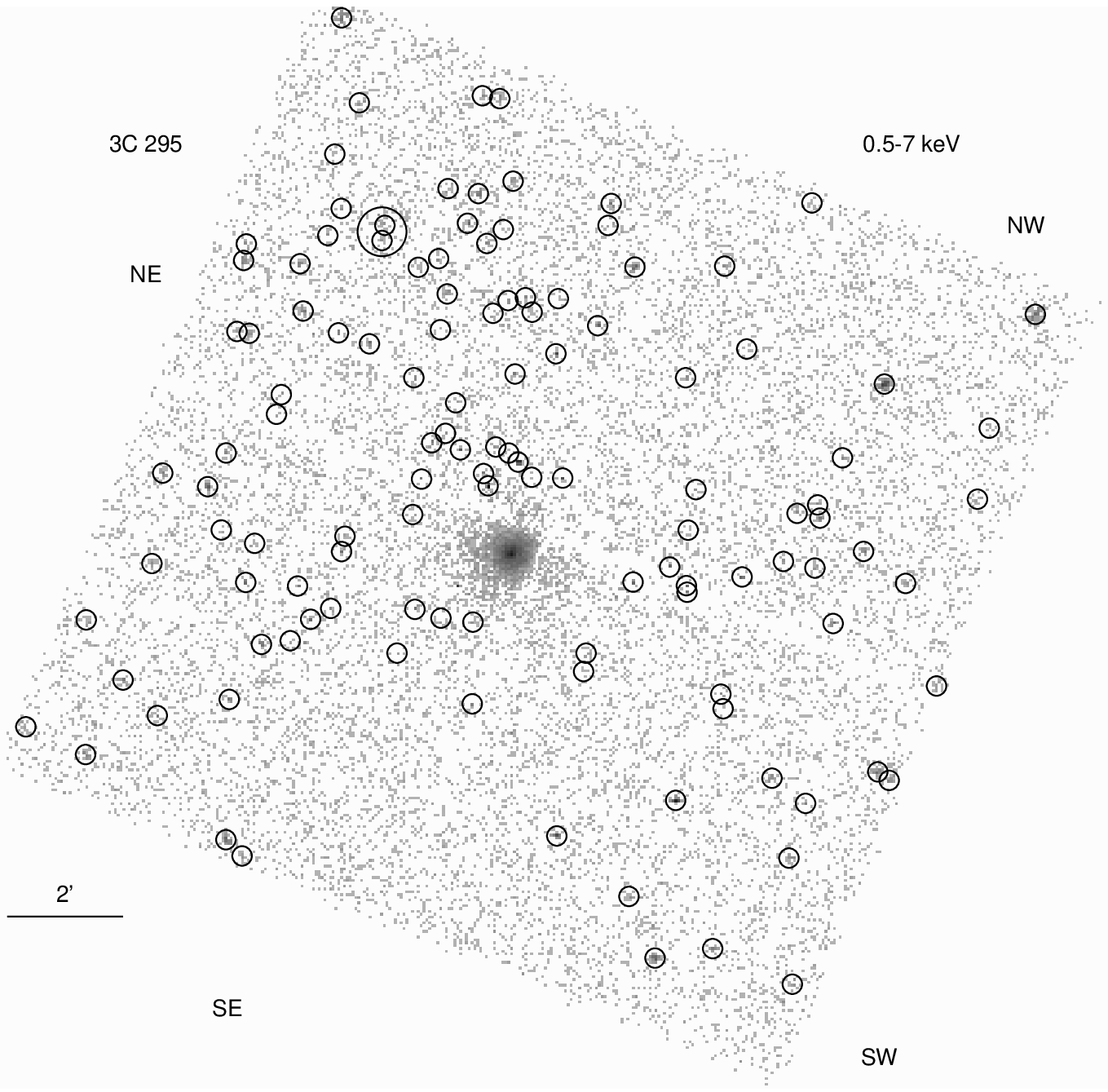} \includegraphics[clip=true]{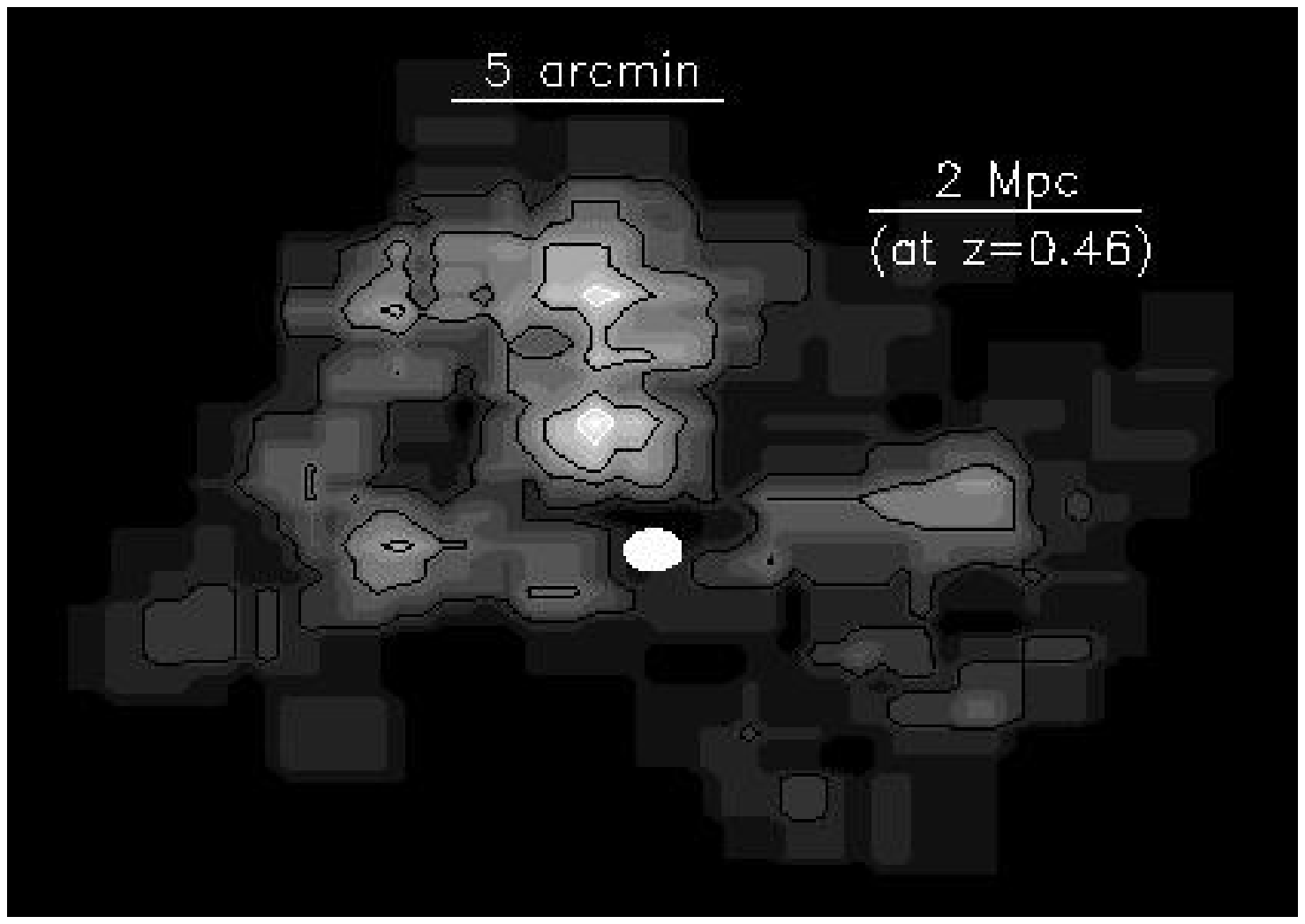}
\includegraphics[clip=true]{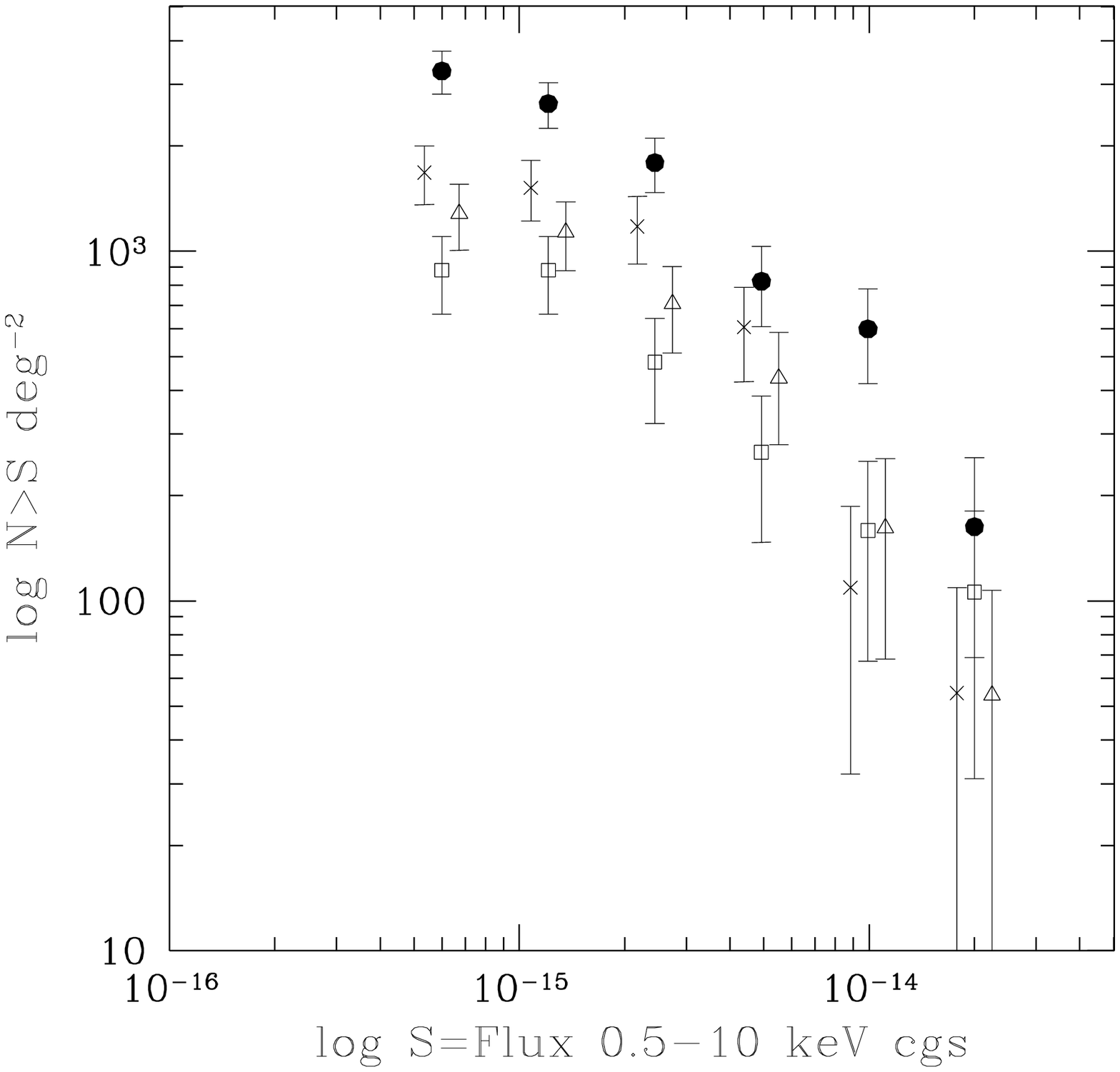}}
     \caption{\footnotesize{Left panel:
 The {\it Chandra} 3C 295 field in the $0.5-7$ keV band. 
Circles represent the sources detected. 
The brightest source in the center of the field
is the cluster of galaxies 3C 295. Central panel: The density profile in the same band.
The linear smoothing factor is $1.5$ arcmin and the four contour levels indicate
source densities of $1.3$, $1.8$, $2.2$ and $2.7$ sources per arcmin$^2$. Right panel: 
The mean 3C 295 LogN LogS in the  $0.5-10$ keV band, calculated for each ACIS-I chip
separately. Filled circles represent counts for the NE chip, open triangles for the
NW, open square for the SW and crosses for the SE.          
               }  } 
    \end{figure}
%

%
%

\section {Observation and Data reduction}

Chandra observed the field around the 
3C 295 cluster with ACIS-I on May 18, 2001, for $92$ ks. 
The data reduction was carried out using the Chandra Interactive
Analysis of Observations software version 2.1.3. 
All the analysis has been performed in the $0.5-2$,
$2-7$ and $0.5-7$ keV bands. 
An identical analysis was performed for the Chandra Deep Field South
(CDFS) in the $0.5-2$ keV and $2-7$ keV bands for comparison, and to check
our analysis methods. 
The source detection was carried out using the `PWDetect' algorithm
(Damiani et al. 1997). We identified $89$ sources
in the $0.5 -2$ keV band, $71$ sources in the $2 -7$ keV band and $121$
sources in the $0.5 -7$ keV band (see fig. 1, left). 
The counts in the $0.5-2$ keV, $2-7$ keV and $0.5-7$ keV
bands were converted in $0.5 -2$ keV, $2 -10$ keV and $0.5 -10$ keV fluxes
using conversion factors appropriate for a $\gamma = 1.8$ power law spectrum
with a galactic absorption toward the 3C 295 field of 
$N_H = 1.33 \times 10^{20}$ cm$^{-2}$. Such values take into account
the quantum efficiency degradation of the CCD.

\section {Analysis}
The following analysis has been perfomed.
The sky coverage has been computed for 3C 295 and CDFS fields.
The LogN-LogS in the soft and hard bands for the whole filelds have
been produced.
The LogN-LogS in the soft, hard and whole bands have been produced
separately for each ACIS-I chip of the 3C 295 observation (see fig. 1, right
for the $0.5-10$ keV band).
3C 295 chip-to-chip and CDFS LogN-LogS have been fitted to evaluate
slopes and normalizations (fig. 2, left for the $0.5-10$ keV band).
A two dimensional Kolmogorov-Smirnov test has been applied to check
whether the sources were uniformly distributed or not.
The angular correlation function (ACF) has been computed in order to
estimate at which scales the sources in the two fields were clustered. 
The ACF has been evaluated also for 
a sample of $8$ Chandra fields with an axposure
time similar to that of 3C 295. 
The errors associated to the functions (fig. 2, centre and right) have been 
calculated using both poisson and bootstrap statistics 
(Barrow, Bhavsar \& Sonoda 1984)

\section{Main results}
The following results have been achieved.
3C 295 and CDFS LogN-LogS are in very good agreement both in the soft
and hard band, and with the CDFS LogN-LogS by Rosati et al. 2002.
The 3C 295 LogN-LogS in the soft, hard and broad bands computed
separately for each ACIS-I chip show an overdensity of sources in the
North-East (NE) chip (fig. 1, right) which reflects the clustering of sources
clearly visible in fig. 1 (left and centre).
The discrepancy between the normalization of the LogN-LogS for the NE
and SW chip is $3.2\;\sigma$, $3.3\;\sigma$ and $4.0\;\sigma$ in the soft, 
hard and broad band, respectively; fig. 2, left shows the 
normalization vs. slope plot for the broad band. 

The two dimensional Kolmogorov-Smirnov test shows that there is a
considerable probability that CDFS sources are uniformly distributed
($P \sim 15\%$ in the soft and hard bands), while the probability that the
3C 295 sources are uniformly distributed is only a few per cent, and
drops below 1\% if we consider the $0.5 - 7$ keV band.  

The angular correlation function of the 3C 295 sources features a strong
signal on scales of a few arcmins, and
also on lower scales in the $0.5 - 7$ keV band (fig. 2, centre). Moreover, the
function is above the value found by Vikhlinin \& Forman (1995) for a
large sample of ROSAT sources. On the other hand, no signs of a
similar behavior is featured by the CDFS and the sample of the $8$ 
Chandra fields (fig. 2, right)

\begin{figure}
\centering
\resizebox{\hsize}{!}{\includegraphics[clip=true]{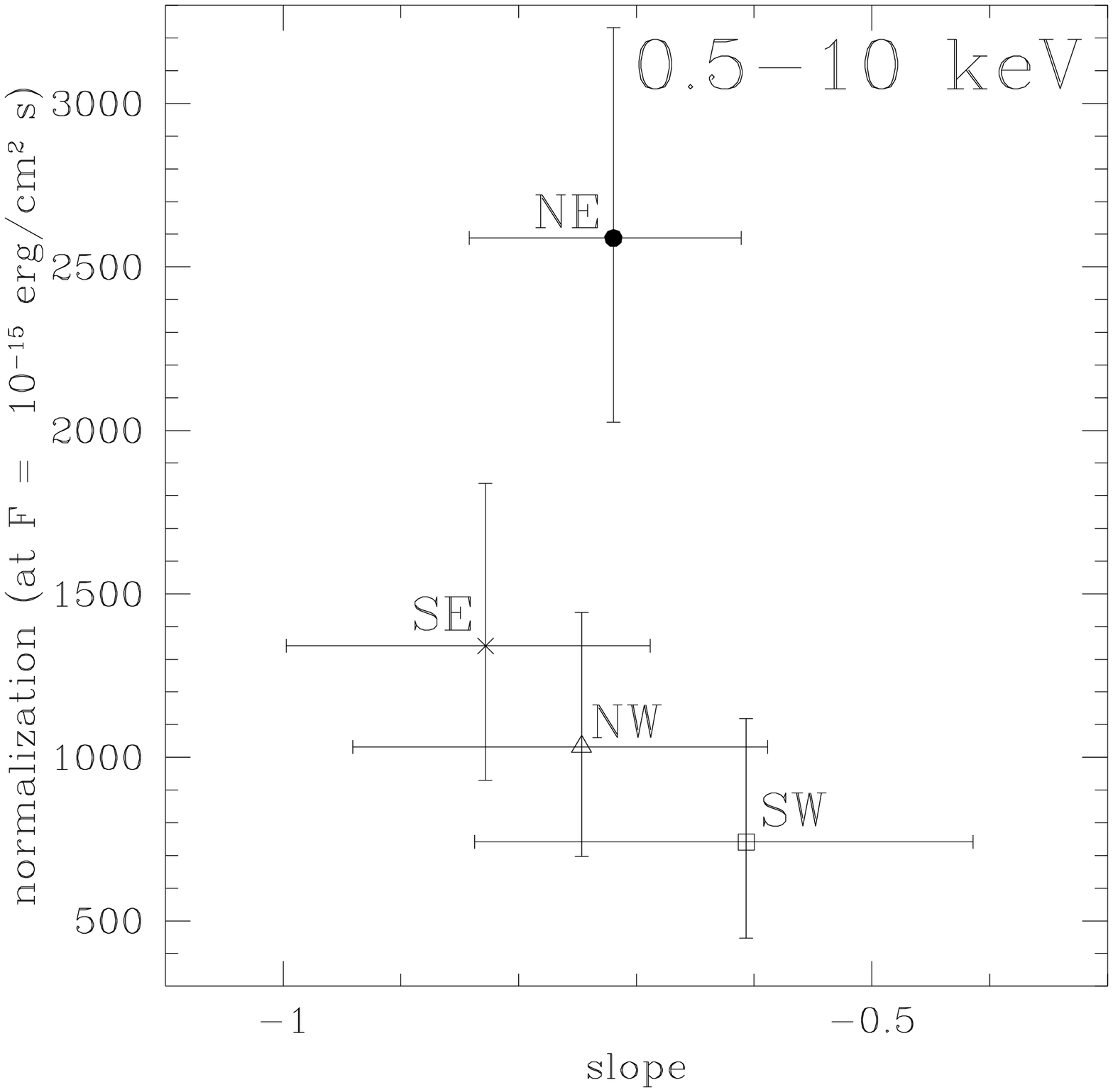}
\includegraphics[clip=true]{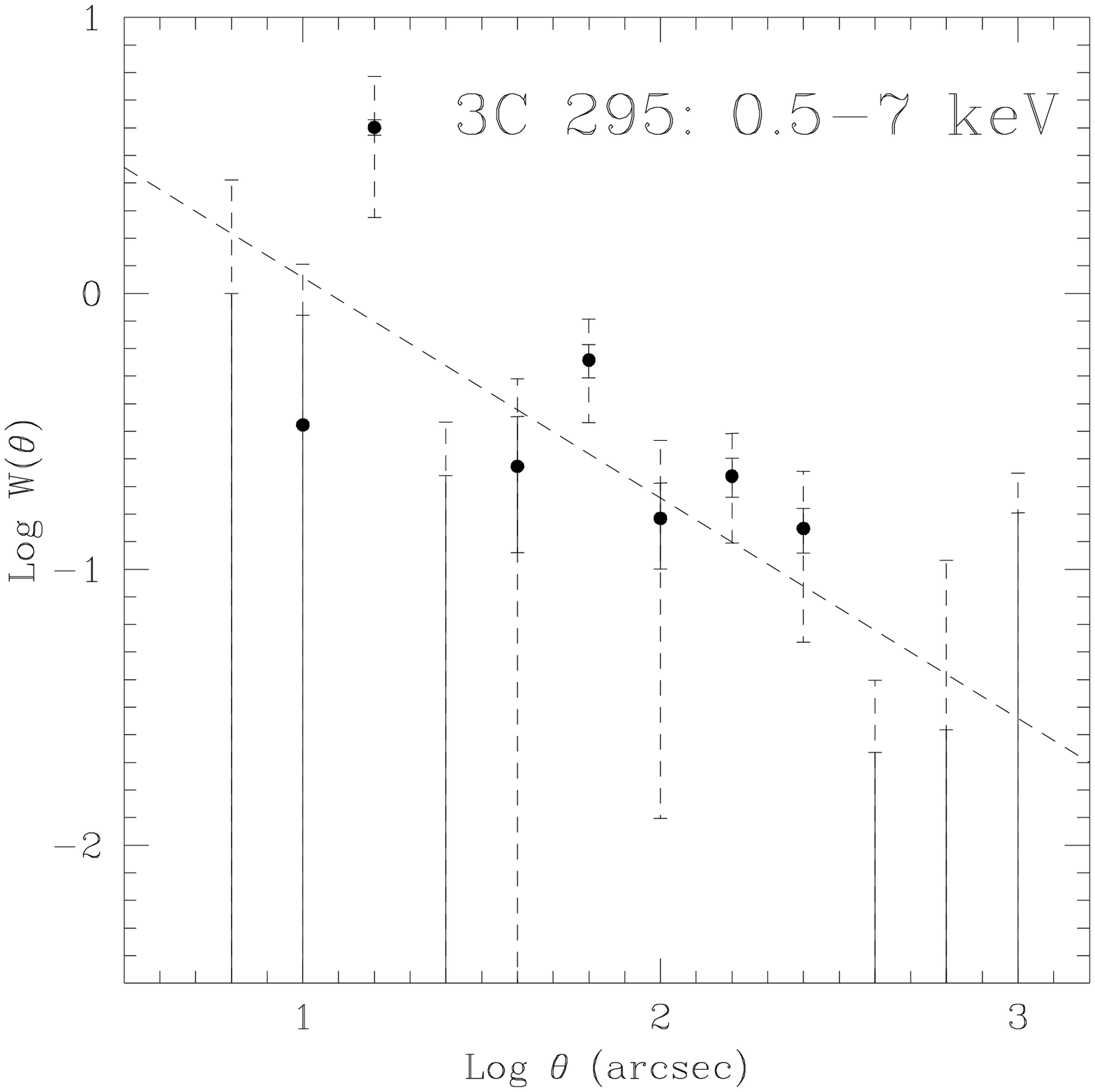}
\includegraphics[clip=true]{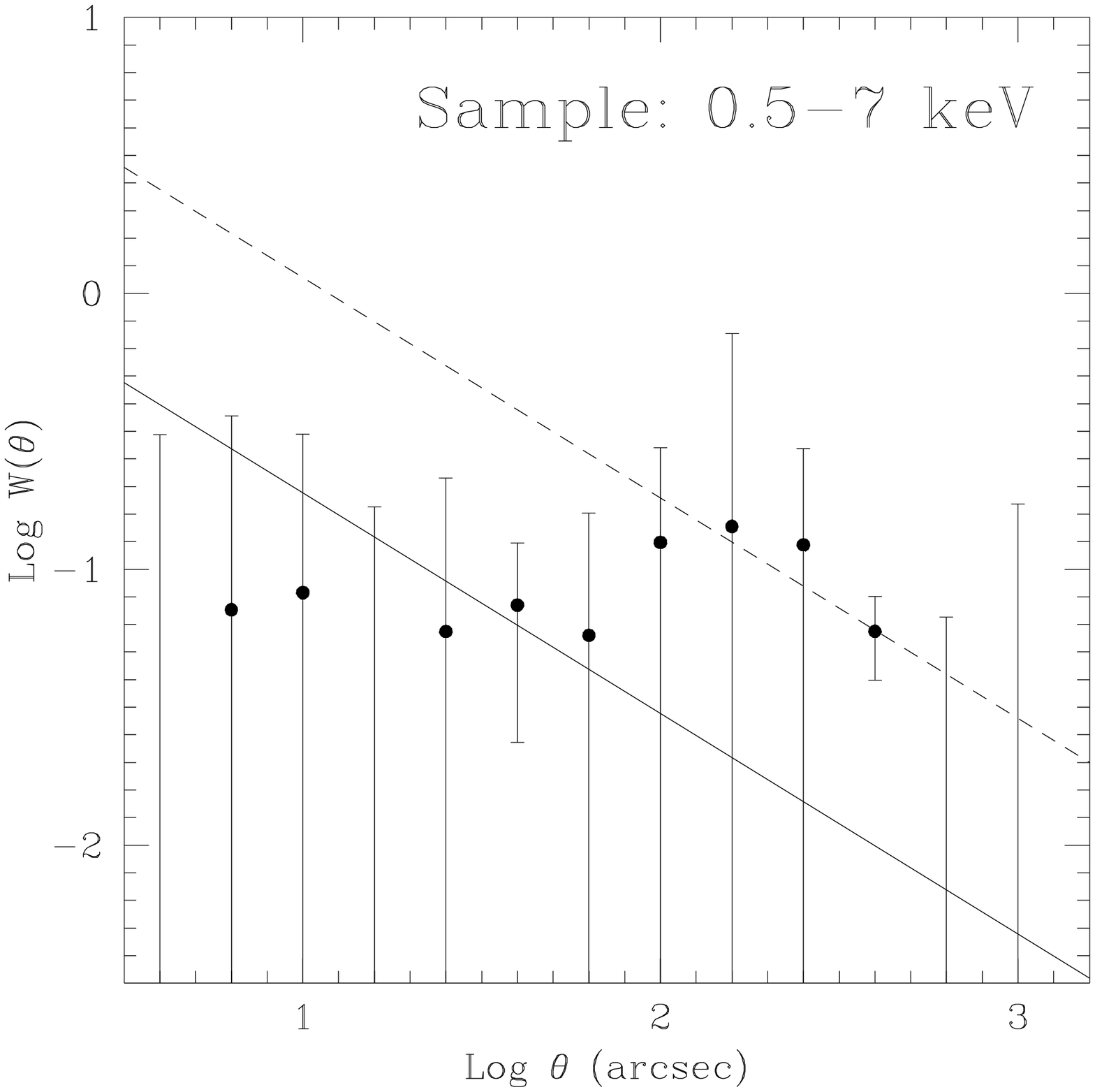}}
\caption{\footnotesize{Left panel: Results of the power law fits to the four LogN-LogS chips
in the broad band. x axis plots the slope of the power law, y axis 
the normalization. Simbles refer to chips as in right panel of fig. 1. Errors are the 
90\% confidence limit. Central panel:
The 3C 295 angular correlation function (ACF)
in the $0.5-7$ kev Band. The dashed line represents the best fit to the 
data. Solid error bars are Poisson; dashed error bars are bootstrap. Left panel: 
The ACF of the sample of Chandra fields in the $0.5-7$ kev Band. The solid
line is the best fit for these data; the dashed line is the fit for the ACF 
of 3C 295 (central panel). Error bars are bootstrap.  }   }
\end{figure}

   \begin{figure}
   \centering
\resizebox{\hsize}{!}{\includegraphics[clip=true,angle=-90]{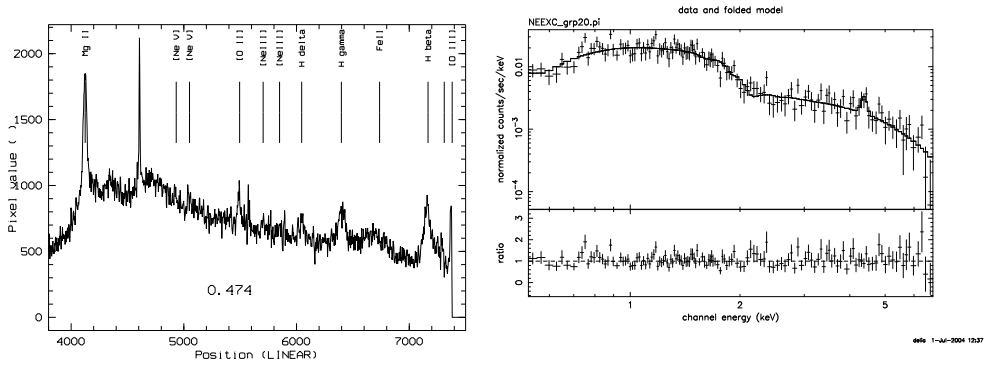} 
\includegraphics[clip=true,angle=-90]{Delia.Fig8.ps}
}
     \caption{\footnotesize{Left panel: The optical spectrum of one of the 
X-ray source in the NE chip, lying close to the cluster.
Right panel: The stacked X-ray spectrum of all the NE sources.             
               }  } 
    \end{figure}

\section{Association with the cluster: first hints}
A work to verify if the overdensity is associated to 3C 295 is in progress. 
Two nights of observation at the TNG have been obtained, but unfortunately 
only a few hours of observing time were completed due to technical and 
weather problems. For this reason we could spectroscopically identify 
only ten sources. Nevertheless, we identified a $z=0.47$ source in the NE 
chip close to the cluster (see fig. 3, left); moreover, we found three 
more sources with a redshift in the range $0.37 \div 0.53$. Although more 
optical identifications are needed, this is a first
clue that points in the direction of on association between the cluster and 
the overdensity. Waiting for more optical data to come,
we searched for more evidences of this association. We extracted the stacked
spectrum of all the NE sources and we fitted it with a power law plus a double
warm absorber ($z=0$ and $z=0.46$) model. Then we added a line to this model
at $E=4.38$ keV, the Energy of the Fe line at $E=6.4 keV$ for a 
source at $z=0.46$ (see fig. 3, right); the goodness of the model 
increased and the normalization of the line is $8^{+12} _{-6} \times 10^{-7 }$
counts s$^{-1}$ keV$^{-1}$ at the $2 \sigma$ confidence level. 
This is another clue for the association we are seeking.

\section{Conclusions}

In this work we studied the excess of sources clearly visible in the upper 
left corner of the Chandra observation of the 3C 295 galaxy cluster field 
(see fig. 1, see D'Elia et al. 2004 for more details). Since N-body and 
hydrodynamical simulations show that clusters of galaxies lie at the nexus of 
several filaments, this excess could represent a filament of the large scale
structure of the Universe.

Moreover, if the redshift of our sources were 
the same of 3C 295 ($z=0.46$) the galaxy overdensity of the field 
(assuming a spherical distribution) is intriguingly close to the expected
galaxy overdensity of filaments.
Works are in progress to check this possibility (D'Elia et al. in prep). 
A first set of optical data have been analyzed and a few sources close to 
the central cluster have been identified. In addition, the X-ray stacked 
spectrum of the excess sources shows an emission line at $E=4.38$ keV 
($2\sigma$ confidence level) which could be the redshifted 
Fe line at $E=6.4$ keV. More optical and new X-ray data are needed in
order to confirm the association and to study the filament properties
to larger scales.


\bibliographystyle{aa}

\begin{thebibliography}{}

\bibitem[]{} Barrow, J. D.,  Bhavsar, S. P.,  \&  Sonoda, D. H., 1984, MNRAS, 210, 19
\bibitem[]{} Cappi, M. et al., 2001, ApJ, 548, 624
\bibitem[]{} Damiani, F., Maggio, A.,  Micela, G. \& Sciortino, S., 1997, ApJ, 483, 350
\bibitem[]{} D'Elia, V. et al., 2004, A\&A in press (astro-ph/0403401)
\bibitem[]{} Hasinger, G. et al., 1998, A\&A, 329, 482
\bibitem[]{} Peacock, J. A., 1999, ``Cosmological Physics'' (Cambridge: Cambridge University Press)
\bibitem[]{} Rosati, P. et al., 2002, ApJ, 566, 667
\bibitem[]{} Vikhlinin, A. \& Forman, A., 1995, ApJL, 455, 109
 

\end{thebibliography}

\end{document}